\documentclass[authoryear,review,3p,times]{elsarticle}

 \usepackage{graphicx}

\usepackage{amsmath, amsthm, amssymb}
\usepackage{natbib}
\usepackage{lineno}
\journal{E\&PSL}

\begin{document}

\begin{frontmatter}



\title{A new model of cosmogenic production of radiocarbon $^{14}$C in the atmosphere}


\author[1]{Gennady A. Kovaltsov}
\author[2,3]{Alexander Mishev}
\author[2,4]{Ilya G. Usoskin\corref{cor}}
\address[1]{Ioffe Physical-Technical Institute, St.Petersburg, Russia}
\address[2]{Sodankyl\"a Geophysical Observatory (Oulu unit), University of Oulu, Finland}
\address[3]{Institute for Nuclear Research and Nuclear Energy, Bulgarian Academy of Sciences, 72 Tzarigradsko chaussee, 1784 Sofia, Bulgaria}
\address[4]{Dept. of Physical Sciences, University of Oulu, Finland}
\cortext[cor]{Corresponding author, e-mail: ilya.usoskin@oulu.fi\\
 {Authors are listed in the alphabetical order}}


\begin{abstract}
We present the results of full new calculation of radiocarbon $^{14}$C production
 in the Earth atmosphere, using a numerical Monte-Carlo model.
We provide, for the first time, a tabulated $^{14}$C yield function for the energy
 of primary cosmic ray particles ranging from 0.1 to 1000 GeV/nucleon.
We have calculated the global production rate of $^{14}$C, which is 1.64 and 1.88 at/cm$^2$/sec
 for the modern time and for the pre-industrial epoch, respectively.
This is close to the values obtained from the carbon cycle reservoir inventory.
We argue that earlier models overestimated the global $^{14}$C production rate
 because of outdated spectra of cosmic ray heavier nuclei.
The mean contribution of solar energetic particles to the global $^{14}$C is calculated
 as about 0.25\% for the modern epoch.
Our model provides a new tool to calculate the $^{14}$C production in the Earth's atmosphere,
 which can be applied, e.g., to reconstructions of solar activity in the past.
\end{abstract}


\end{frontmatter}

\linenumbers

\section{Introduction}

Radiocarbon $^{14}$C is a long-living (half-life about 5730 years) radioactive nuclide produced mostly by cosmic rays in the Earth's atmosphere.
Soon after production, it gets oxidized to $^{14}$CO$_2$ and in the gaseous form takes part in
 the complex global carbon cycle \citep{bolin79}.
Radiocarbon is important not only because it is used for dating in many applications \citep[e.g., ][]{dorman04,kromer09},
 but also because it forms a primary method of paleo-reconstructions of solar activity on the
 millennial time scales \citep[e.g.,][]{stuiver80,stuiver89,bard97,muscheler07}.
An essential part of the solar activity reconstruction from radiocarbon data is computation of $^{14}$C production by cosmic rays in the
 Earth's atmosphere.
First such computations were performed in the 1960--1970s \citep[e.g.,][]{lingenfelter63,lingenfelter70,light73,obrien79}
 and were based on simplified numerical or semi-empirical methods.
Later, full Monte-Carlo simulations of the cosmic-ray induced atmospheric cascade had been performed
 \citep{masarik99,masarik09}.
Most of earlier models, including \cite{obrien79} and \citet{masarik99} deal with a prescribed functional shape of the galactic cosmic ray spectrum, which
 makes it impossible to be applied to other types of cosmic ray spectra, e.g., solar energetic particles,
 supernova explosions, etc.
A flexible approach includes calculation of the yield function (the number of cosmogenic nuclei produced in
 the atmosphere by the primary cosmic rays of the given type with the fixed energy and unit intensity outside the atmosphere),
 which can be convoluted with any given energy spectrum of the primary cosmic rays
 \citep[e.g.,][]{webber03,webber07,usoskin_7Be_08,kovaltsov_Be10_10}.
This approach can be directly applied to, e.g., a problem of the signatures of extreme solar energetic particle
 events in the cosmogenic nuclide data, which is actively discussed \citep[e.g.,][]{usoskin_GRL_SCR06,hudson10,lavaletta11}.
Some earlier models \citep{lingenfelter63,castagnoli80} provide the $^{14}$C yield function however it is limited
 in energy.
Moreover, different models give results which differ by up to 50\% from each other, leading to large uncertainty in the global $^{14}$C production rate.
Therefore, the present status is that models providing the yield function are 30--50 years old {\bf and have large uncertainties}.

In addition, there is a systematic discrepancy between the results of theoretical models for the $^{14}$C production and
 the global average $^{14}$C production rate obtained from direct measurements of the specific
 $^{14}$CO$_2$ activity in the atmosphere and from the carbon cycle reservoir inventory.
While earlier production models predict that the global average pre-industrial production rate should be 1.9--2.5 atoms/cm$^2$/sec,
 estimates from the carbon cycle inventory give systematically lower values ranging between 1.6 and 1.8 atoms/cm$^2$/sec
 \citep{lingenfelter63,lal68,damon89,obrien91,goslar01,dorman04}.
This discrepancy is known since long \citep{lingenfelter63} but is yet unresolved \citep{goslar01}.

In this work we redo all the detailed Monte-Carlo computations of the cosmic-ray induced atmospheric cascade
 and the production of $^{14}$C in the atmosphere to resolve the problems mentioned above.
In Section \ref{S:model} we describe the numerical model and calculation of the radiocarbon production.
In Section \ref{S:comp} we compare the obtained results with earlier models.
In Section \ref{S:SEP} we apply the model to calculate the $^{14}$C production by galactic cosmic rays and solar energetic
 particle events for the last solar cycle.
Conclusions are presented in Section \ref{S:disc}.

\section{Calculation of the $^{14}$C production}
\label{S:model}

Energetic primary cosmic ray particles, when entering the atmosphere, collide with nuclei of the atmospheric gases
 initiating a complicated nucleonic cascade (also called shower).
Here we are interested primarily in secondary neutrons whose distribution in the atmosphere varies
 with altitude, latitude, atmospheric state and solar activity.
Neutrons are produced in the atmosphere through multiple reactions including high-energy direct reactions, low-energy
 compound nucleus reactions and evaporation of neutrons from the final equilibrium state.
Most of neutrons with energy below 10 MeV are produced as an evaporation product of excited nuclei, while high-energy
 neutrons originate as knock-on neutrons in collisions or in charge exchange reactions of high-energy protons.
While knock-on neutrons are mainly emitted in the forward direction (viz. downwards), evaporated neutrons of lower energy are
 nearly isotropic.
Radiocarbon $^{14}$C is a by-product of the nucleonic cascade, with the main channel being
 through capture of secondary neutrons by nitrogen: N14(n,p)C14.
Other channels (e.g., via spallation reactions) contribute negligibly, but are also considered here.

We have performed a full Monte Carlo simulation of the nucleonic component of the cosmic ray induced
 atmospheric cascade, using the Planetocosmic code \citep{desorgher05}
 based on GEANT-4 toolkit for the passage of particles through matter \citep{agostinelli03}
 (see details in Appendix).
The secondary particles were tracked through the atmosphere until they undergo reactions
 with an air nucleus, exit the atmosphere or decay.
In particular, secondary neutrons were traced down to epi-thermal energy.
Simulations are computationally intensive.
Simulations of single energies (ranging from 0.1 to 1000 GeV/nuc) were conducted, to determine the resulting
 flux of secondary neutrons.
Since the calculations require very large computational time to keep the
 statistical significance of the results for low energies, we applied an
 analytical approach for atmospheric neutrons with energy below 10 eV (see details in Appendix).
Cross-sections have been adopted from the Experimental Nuclear Reaction Database (EXFOR/CSISRS)
 http://www.nndc.bnl.gov/exfor/exfor00.htm.
The number of simulated cascades induced by primary CR particles was chosen as $10^5-10^6$ to keep the statistical stability of the results
 at a reasonable computational time.
Computations were carried out separately for primary protons and $\alpha-$particles.
Because of the similar rigidity/energy ratio, nuclei with $Z>2$ were considered as effectively $\alpha-$particles
 with the scaled number of nucleons \citep[cf.][]{usoskin_7Be_08}.

As the main result of these detailed computations we calculated the $^{14}$C yield function.
The yield functions for primary protons and $\alpha-$particles are tabulated in Table~\ref{Tab:Y}
 and shown in Fig.~\ref{Fig:Y} (the energy range above 100 GeV/nuc is not shown).
Note that the yields (per nucleon with the same energy) are identical for protons and $\alpha-$particle,
 viz. an $\alpha-$particle is identical to four protons, at energies above 10 GeV/nuc.
Details of the computations are given in Appendix \ref{App}.
All further calculations are made using these yield functions.


In order to compute the $^{14}$C production $q$ in the atmosphere at a certain place and conditions/time,
 one can use the following method:
\begin{equation}
q(t)=\sum_i{\int_{E_{i\rm c}}^\infty{Y_i(E)\, J_i(E,t)\, dE}},
\label{Eq:Q}
\end{equation}
where $E$ is the particle's kinetic energy per nucleon, $J_i$ is the spectrum of primary particles
 of type $i$ on the top of the atmosphere, $E_{i\rm c}$ in GeV/nucleon is the kinetic energy per nucleon corresponding to
 to the local geomagnetic rigidity cutoff $P_{\rm c}$ in GV.
\begin{equation}
P_{\rm c}={A_i\over Z_i}\sqrt{E_{i\rm c}\, (E_{i\rm c}+2\, E_{\rm r})},
\label{eq:Pc}
\end{equation}
where $E_{\rm r}=0.938$ GeV/nucleon is the proton's rest mass.
Summation is over different types of the primary cosmic ray nuclei with charge $Z_i$ and mass $A_i$ numbers.
The local geomagnetic rigidity cutoff is roughly defined via the geomagnetic latitude $\lambda_{\rm G}$ of the location as
 following \citep{elsasser56}
\begin{equation}
{P_{\rm c}\, [{\rm GV}] = 1.9\cdot M\cdot\cos^4{\lambda_{\rm G}}},
\label{eq:dip}
\end{equation}
where $M$ is the dipole moment in units of $[10^{22}$ A m$^2$] of the Earth's magnetic field.
Although this approximation may slightly $\le 2$\% overestimate the $^{14}$C production \citep{obrien08},
 it is sufficient to study the global cosmic ray flux \citep{dorman09,clem97}.
The global production $Q$ of radiocarbon is defined as the spatial global average of the local
 production $q$ (both quantities give the number of $^{14}$C nuclei produced per second per cm$^2$ of the Earth's surface).
For the isotropic flux of primary particles in the interplanetary space (the level of anisotropy
 for galactic cosmic rays is usually smaller than 1\%)
 the global production can be written as:
\begin{equation}
Q(t)=\sum_i{\int_{0}}^\infty{Y_i(E)\, J_i(E,t)\,(1-f(E))\, dE},
\label{Eq:glob}
\end{equation}
where the function
\begin{equation}
 f(E) =
  \begin{cases}
   \sqrt{1-\sqrt{P(E)/ (1.9 \cdot M)}},  & \text{ if  } P \leq 1.9\cdot M \\
   0,        & \text{ if } P> 1.9\cdot M
  \end{cases}
\end{equation}
corresponds to $\sin(\lambda_{\rm G})$ and accounts for the spatial average with the effect of the geomagnetic cutoff.

Substituting any particular particle spectrum $J_i$ into Eq.~\ref{Eq:glob} one can evaluate the $^{14}$C
  production rate for different populations of cosmic rays, e.g.,
 galactic cosmic rays (GCR), solar energetic particles (SEP), or more exotic sources like a nearby supernova explosion.

First we consider the main source of $^{14}$C, GCR modulated by the solar activity, using the standard
 approach.
The energy spectrum of GCR particles of type $i$ at 1 AU, $J_i$,
 is defined by the local interstellar spectrum (LIS), $J_{\mbox{LIS},i}$, and the modulation
 potential $\phi$ as \citep[see the formalism in][]{usoskin_Phi_05}:
\begin{equation}
J_i(E,\phi)=J_{{\rm LIS,}i}(E+\Phi_i){(E)(E+2E_{\rm r})\over (E+\Phi_i)(E+\Phi_i+2E_{\rm r})},
\label{Eq:ff}
\end{equation}
where $\Phi_i=(eZ_i/A_i)\phi$.
The modulation potential $\phi$ is the variable related to solar activity, that
 parameterizes the shape of the modulated GCR spectrum.
The fixed function $J_{\rm LIS}(T)$ is not exactly known and may affect the absolute value of $\phi$ \citep[e.g.,][]{usoskin_Phi_05,webber09,herbst10,obrien10}.
Thus, the exact model of LIS must be specified together with the values of $\phi$.
Here we use, as earlier, the proton LIS in the form \citep{burger00,usoskin_Phi_05}:
 \begin{equation}
J_{\rm LIS}(E)={1.9\times 10^4\cdot P(E)^{-2.78}\over 1+0.4866\, P(E)^{-2.51}},
\label{Eq:LIS}
\end{equation}
 where $P(E)=\sqrt{E(E+2\, E_r)}$, $J$ and $E$ are expressed in units of particles/(m$^2$ sr s GeV/nucleon) and
 in GeV/nucleon, respectively.
Here we consider two species of GCR separately: protons and heavier species, the latter including all particles with $Z>1$ as
 $\alpha-$particles with $Z/A=0.5$ scaled by the number of nucleons.
Heavier species should be treated separately as they are modulated in the
 heliosphere and Earth's magnetosphere differently, compared to protons because of the
 different Z/A ratio.
Here we consider the nucleonic ratio of heavier particles (including $\alpha-$particles) to protons in
 the interstellar medium as 0.3 \citep{webber03,gaisser10}.

\begin{figure}
\begin{center}
\resizebox{8cm}{!}{\includegraphics{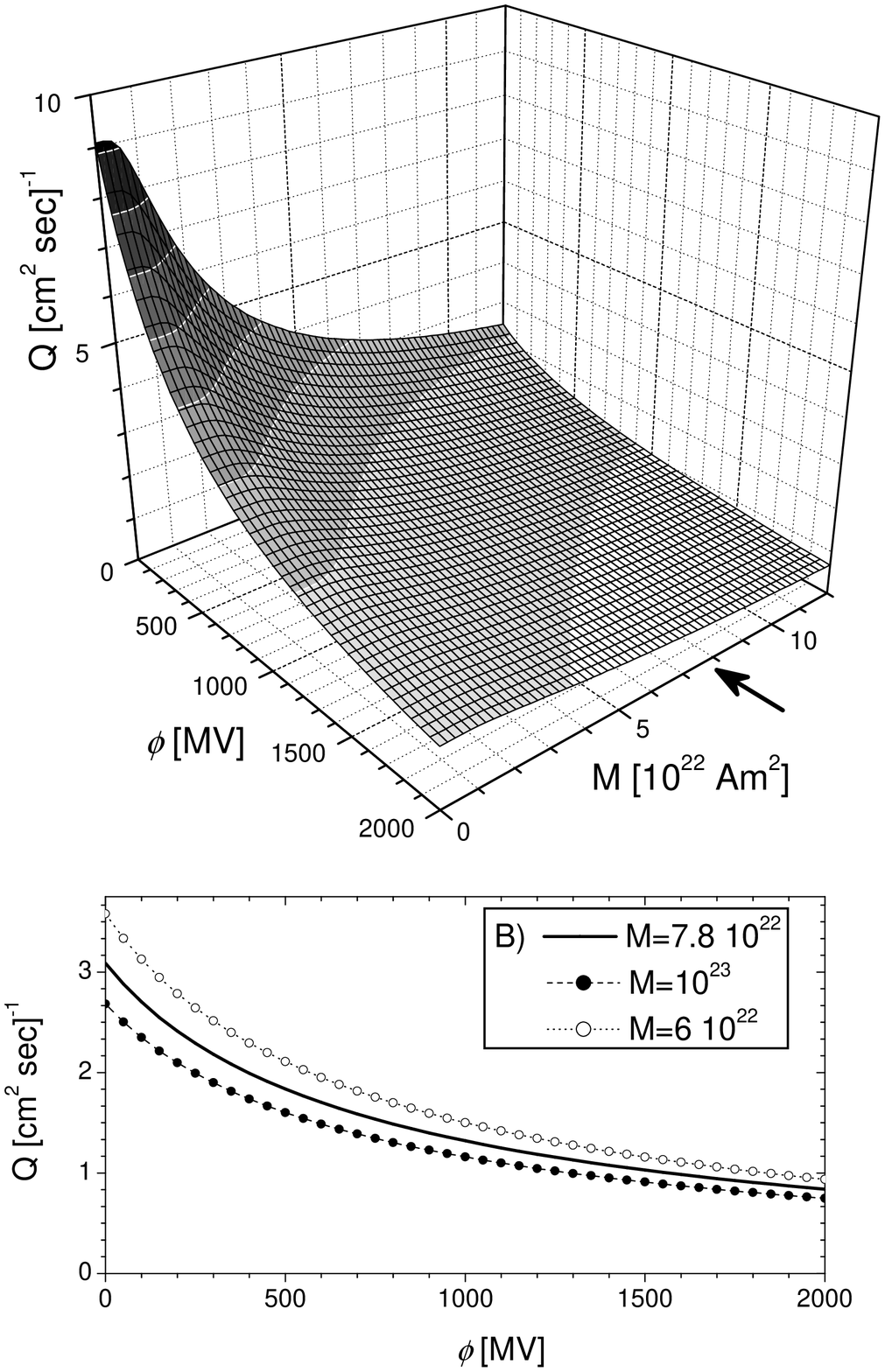}}
\end{center}
\caption{
The global production of $^{14}$C as function of the modulation potential $\phi$ and the geomagnetic
 dipole moment $M$.
 The present value of $M=7.8\cdot 10^{22}$ A m$^2$ is indicated by the thick arrow.
 The lower panel shows three cross-sections of the upper panel corresponding to the
  present value as well as to the maximum and minimum values of $M$ over the past millennia,
  as indicated in the legend.
  Digital table for this plot is available at electronic supplement for this paper.
\label{Fig:prostynya}}
\end{figure}

The global $^{14}$C production $Q$ by GCR depends on two parameters, the solar magnetic activity quantified
 via the modulation potential $\phi$ and the Earth's geomagnetic field (its dipole moment $M$).
The dependence is shown in the upper panel of Fig.~\ref{Fig:prostynya}.
One can see that both parameters are equally important, and the knowledge of the geomagnetic field
 is very important \citep{snowball07}.
In the lower panel, three cuts of the upper panel are shown to illustrate the effect of solar activity on $Q$, for the fixed
 geomagnetic field, corresponding to the modern conditions $M=7.8\cdot10^{22}$ A m$^2$, as well as
 maximum ($10^{23}$ A m$^2$) and minimum ($6\cdot 10^{22}$ A m$^2$) dipole strength over the last ten millennia
 of the Holocene \citep{korte11}.
The response of $Q$ to changes of the geomagnetic field during the Holocene is within $\pm 15$\%.
However, the global $^{14}$C would be nearly doubled during an inversion of the geomagnetic field (viz. $M\rightarrow 0$).
The modulation potential $\phi$ varies between about 300 and 1500 MV within a modern high solar cycle \citep{usoskin_bazi_11},
 and can be as low as about 100 MV during the Maunder minimum \citep{mccracken04,usoskin_AA_07,steinhilber08}.
Thus, changes of the solar modulation can also lead to a factor of 2--3 variability on the global
 $^{14}$C production rate.

\begin{figure}
\begin{center}
\resizebox{\hsize}{!}{\includegraphics{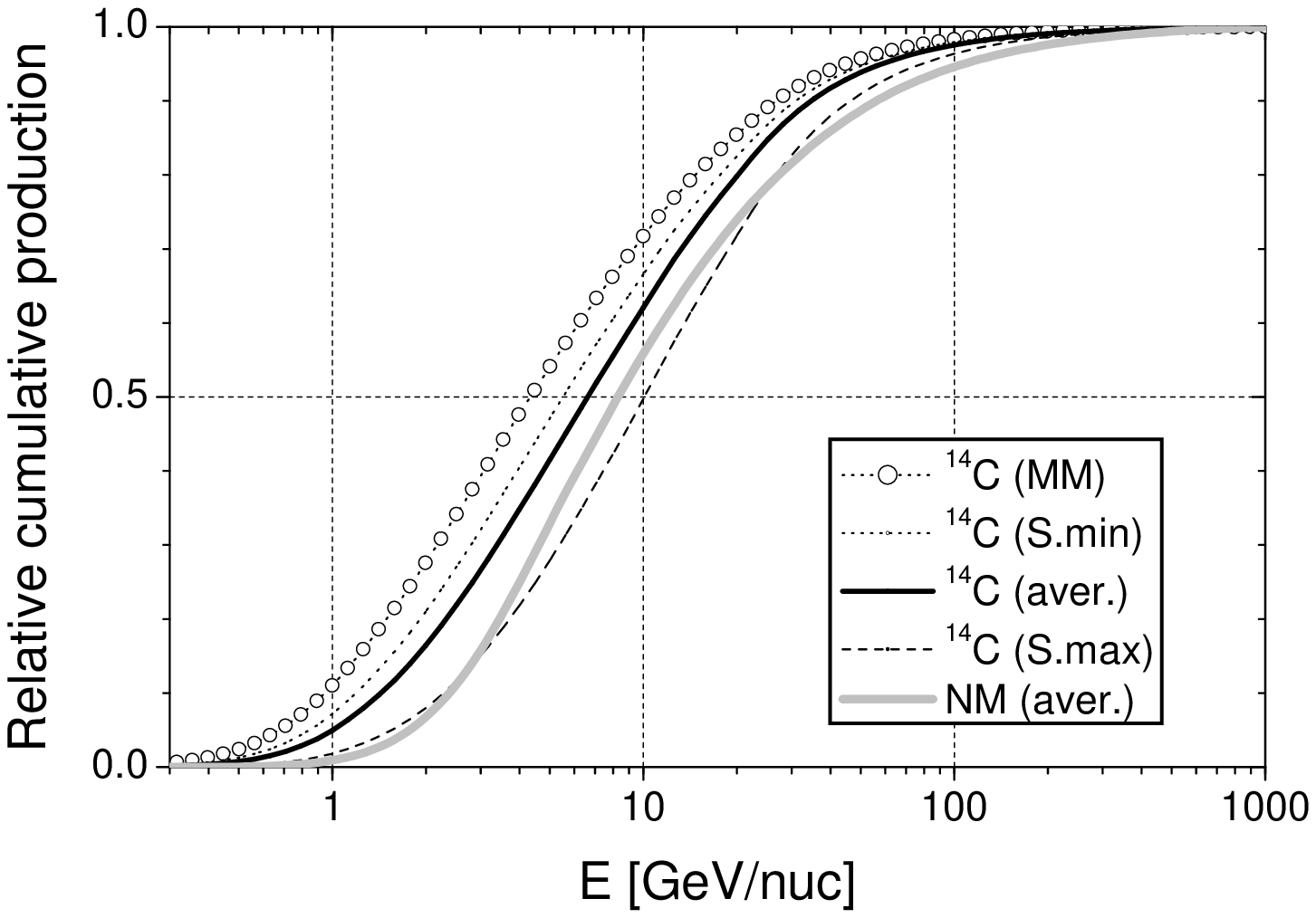}}
\end{center}
\caption{
Relative cumulative production of $^{14}$C (fraction of the total production)
 as a function of the primary cosmic ray energy for different conditions:
 average solar activity (solid "aver." curve), solar maximum ($\phi=1200$ MV, dashed "S.max" curve),
 solar minimum ($\phi=300$ MV, dotted "S.min" curve), Maunder minimum ($\phi=100$ MV, circled "MM" curve).
 The thick grey curve corresponds to a polar sea-level neutron monitor.
 All curves are shown for the modern Earth magnetic field.
\label{Fig:dQ}}
\end{figure}

Next we investigated the sensitivity of $Q$ to the energy of GCR.
In Fig.~\ref{Fig:dQ} we show the relative cumulative production of $^{14}$C, viz. the fraction of the
 total production caused by primary cosmic rays with energy below the given value $E$, as a function of $E$
 for different conditions.
Often the median energy (the energy which halves the production) is used as a characteristic energy \citep[e.g.,][]{lockwood96},
 which is the crossing of the curves in Fig.~\ref{Fig:dQ} with the horizontal dashed line.
One can see that the median energy of $^{14}$C production slightly changes with the level of solar activity,
 varying between 4 and 10 GeV/nuc corresponding to the Maunder minimum and the maximum of a strong solar cycle, respectively.
The sensitivity of $Q$ to the energy of GCR is close to that of a sea-level polar neutron monitor \citep[cf.][]{beer_NM_00}.
Slightly different shape of the neutron monitor cumulative response is due to the fact that it is
 ground-based while $^{14}$C is produced in the entire atmosphere.

\begin{figure}
\begin{center}
\resizebox{\hsize}{!}{\includegraphics{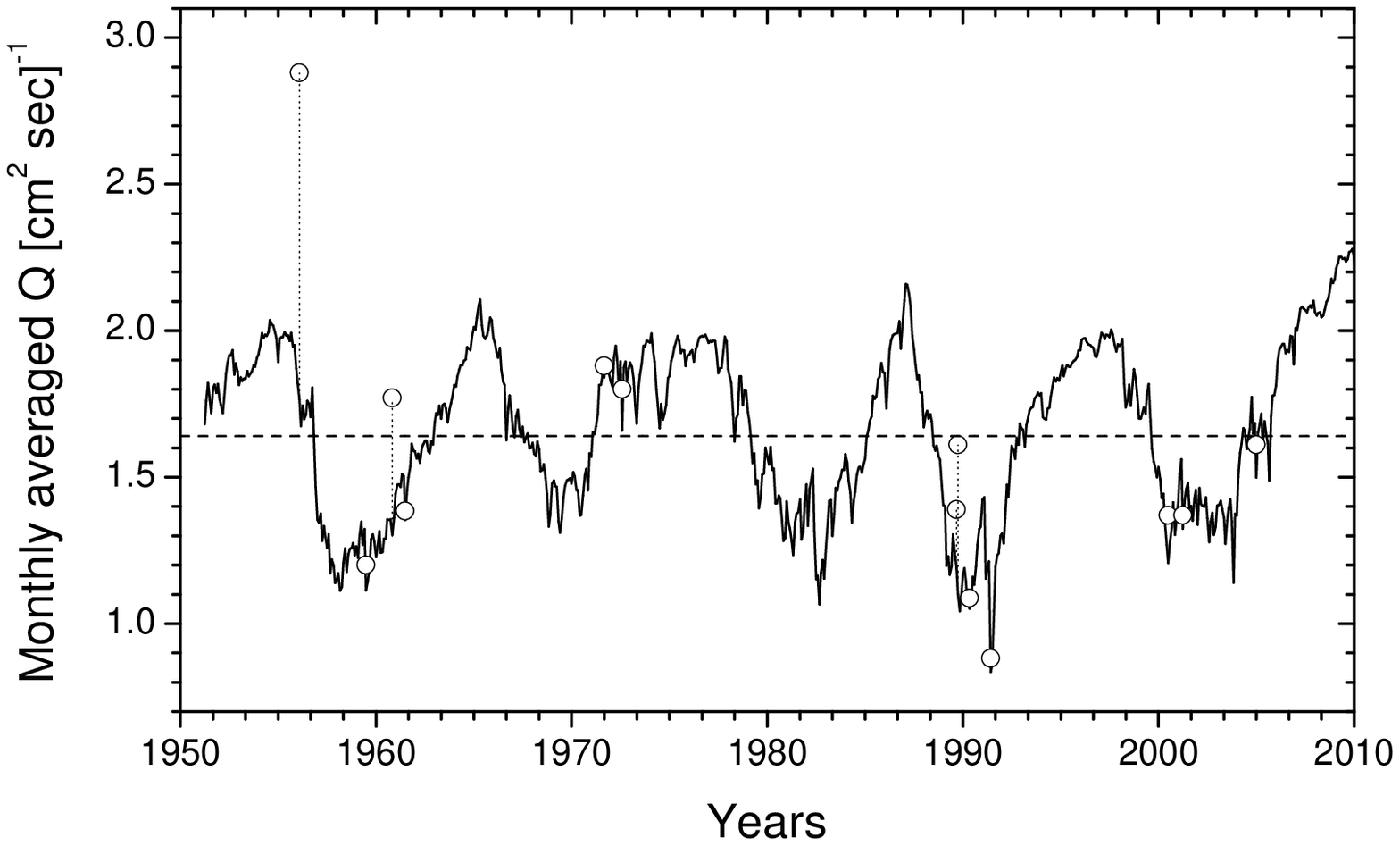}}
\end{center}
\caption{
Monthly averaged global production rates of $^{14}$C since 1951 calculated using cosmic rays data from the world
 network of neutron monitors and our calculated yield function.
Open circles correspond to months with major solar energetic particle events.
\label{Fig:50y}}
\end{figure}

As an example, we calculated the $^{14}$C production predicted by the model for the last 60 years
 (see Fig.~\ref{Fig:50y}) using the GCR modulation, reconstructed from the ground-based network
 of neutron monitors \citep{usoskin_bazi_11}, and IGRF (International Geomagnetic Reference Field --
 http://www.ngdc.noaa.gov/IAGA/vmod/igrf.html) model of the Earth's magnetic field.
The mean radiocarbon production for that period (1951--2010) is $Q=1.64$ atom/cm$^2$/sec,
 with the variability by a factor of two between 1.1 (in 1990 solar maximum) and 2.2 (in 2010 solar minimum) atom/cm$^2$/sec.

The mean $^{14}$C production for the pre-industrial period (1750--1900) calculated using the GCR modulation
 reconstruction by \citet{alanko07} and paleomagnetic data by \citet{korte11} is 1.88 atom/cm$^2$/sec which is
 essentially lower than those reported in earlier works (1.9--2.5 atom/cm$^2$/sec) and closer to the
 values obtained from the carbon cycle inventory (1.6--1.8 atom/cm$^2$/sec) -- see Introduction.
This values can be further $\approx 2$\% lower because of the used geomagnetic cut-off approach \citep{obrien08}.


\section{Comparison with earlier models}
\label{S:comp}

\begin{figure}
\begin{center}
\resizebox{\hsize}{!}{\includegraphics{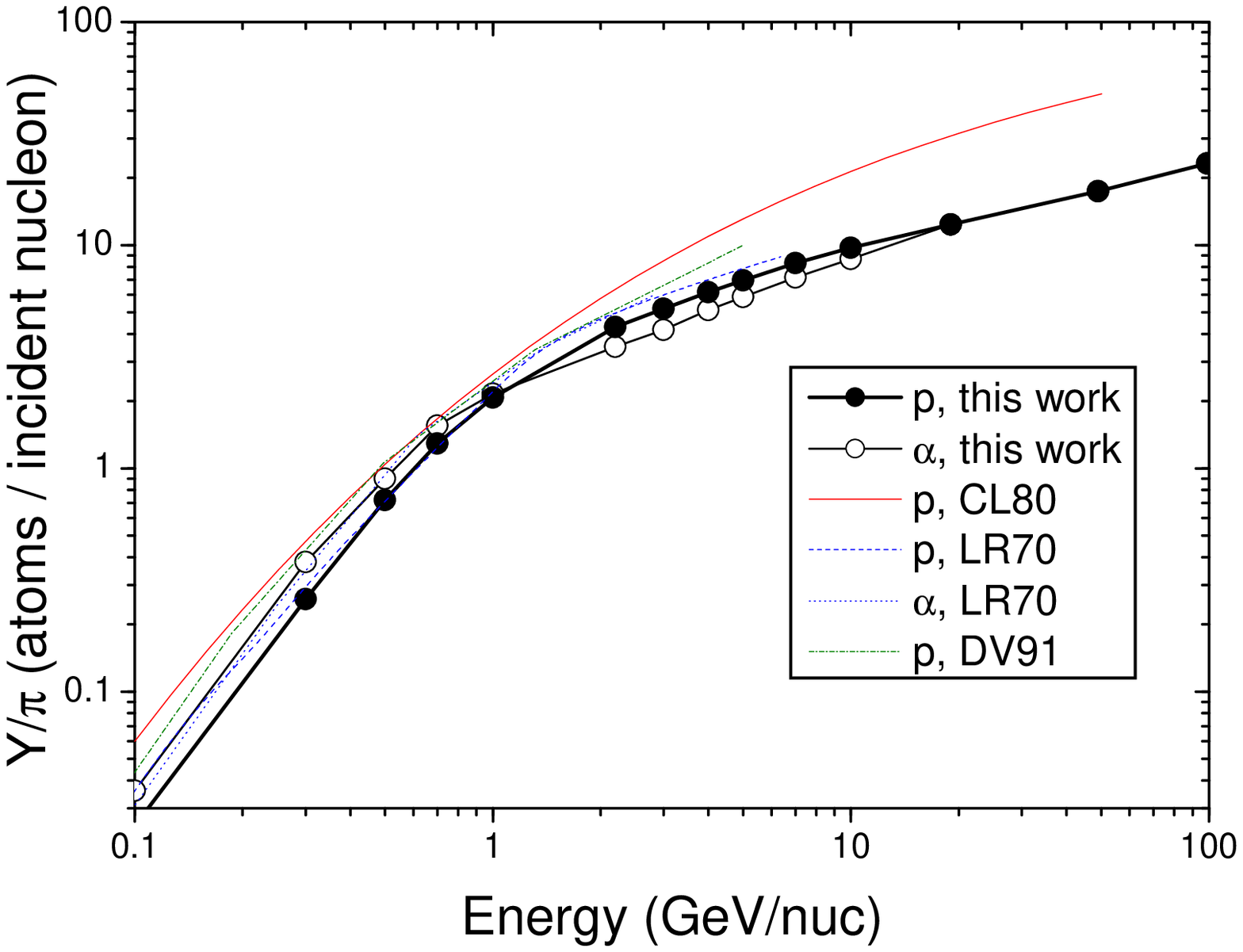}}
\end{center}
\caption{
Yield function $Y/\pi$ of $^{14}$C production in the Earth's atmosphere by primary cosmic
 rays protons and $\alpha-$particles (as denoted by "p" and "$\alpha$" in the legend,
 respectively) with given energy per nucleon.
Different curves correspond to the present work (Table \ref{Tab:Y}) and earlier models \citep[CL80 - ][]{castagnoli80},
 \citep[LR70 -- ][]{lingenfelter70} and \citep[DV91 -- ][]{dergachev91}, as denoted in the legend.
\label{Fig:Y}}
\end{figure}

In Fig.~\ref{Fig:Y} we compare our present results with the yield functions calculated earlier (see
 the Figure caption for references).
Our results are consistent with most of the earlier calculations (LR70 and DV91) within 10-20\%.
The CL80 yield function is not independently calculated but modified from LR70.
While it is formally given for protons it effectively includes also $\alpha-$particles via scaling,
 thus being systematically higher than the other yield functions.
Note that all the earlier computations of the yield function were limited in energy so that
 the upper considered energy of primary cosmic rays was from several to 50 GeV/nuc.
On the other hand, contribution of higher energy cosmic rays is significant and may reach half
 of the total $^{14}$C production (see Fig.~\ref{Fig:dQ}).
Here we present, for the first time, the $^{14}$C yield function calculated up to TeV/nuc energy.
Contribution from the higher energies is negligible because of the steep spectrum of GCR.

\begin{figure}
\begin{center}
\resizebox{\hsize}{!}{\includegraphics{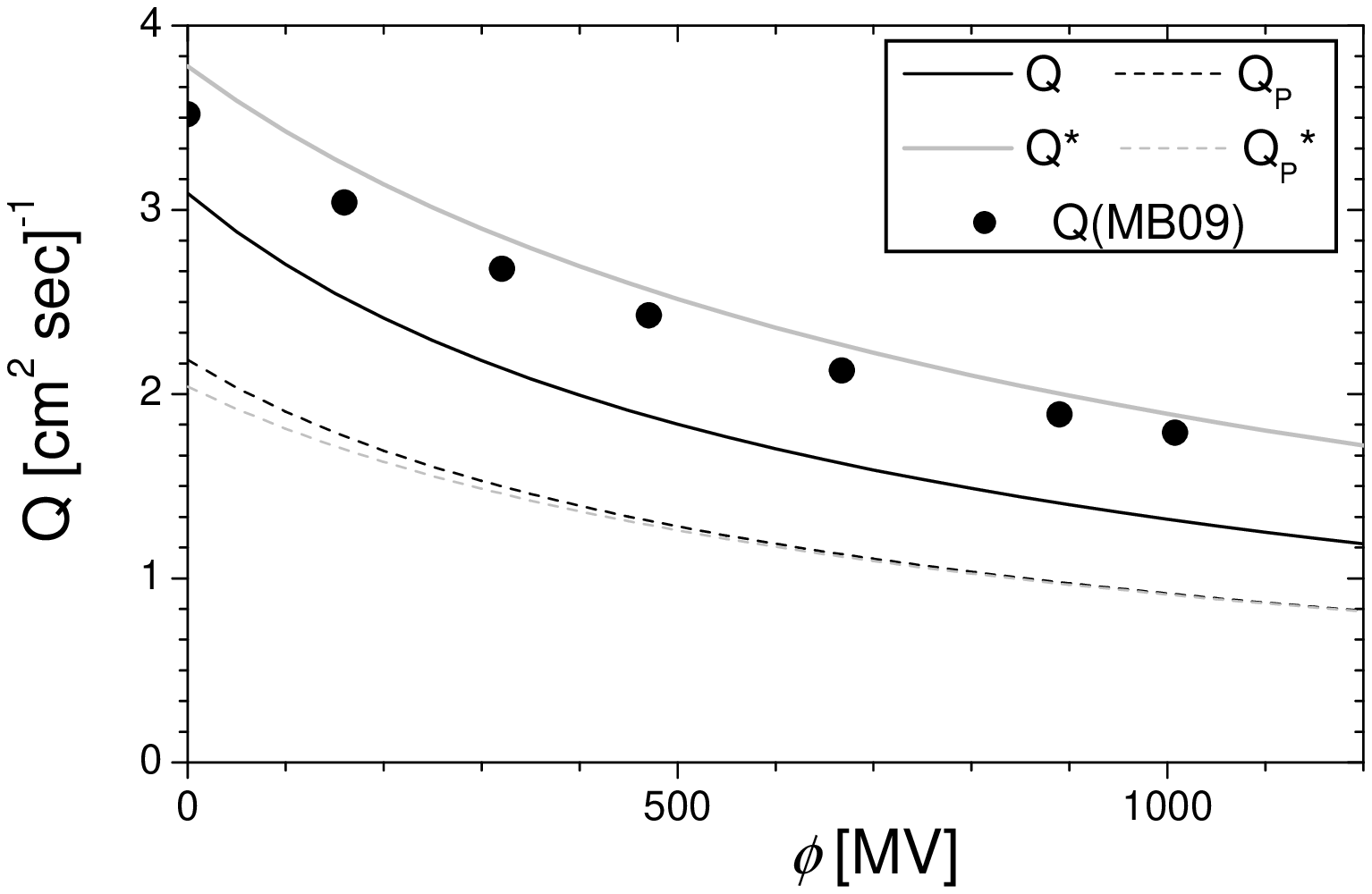}}
\end{center}
\caption{
Comparison of the global $^{14}$C production rates, computed by different models
 as function of the modulation potential for the modern geomagnetic field.
Big dots correspond to the original results by \citet{masarik09}.
Curves are computed using our calculated yield function (Table~\ref{Tab:Y}) and applying
 different cosmic rays spectra.
Black curves (Q values) are calculated using the present results, while grey curves (Q* values)
 are calculated using our yield function but GCR spectra as used by \citet{masarik09}.
Solid and dashed lines correspond to the total production and to production only by primary protons, respectively.
\label{Fig:MB}}
\end{figure}

Next we perform a more detailed comparison with the most recent $^{14}$C production model by \citet[][ - MB09]{masarik09},
 who also used a GEANT-4 Monte-Carlo simulation tool.
Since MB09 did not calculate the yield function, we use another way of comparison, via computing the
 global averaged $^{14}$C production rate, as illustrated in Fig.~\ref{Fig:MB}.
Our present result (black curve $Q$) in the Figure is systematically lower than that given by MB09 (big dots)
 by 25-30\%.
We suspect that the discrepancy arises from that \citet{masarik09} calculated the $^{14}$C production for a
 prescribed GCR spectrum in the form given by \citep{garcia75,castagnoli80}, which is different from the spectrum we
 use here \citep[adopted from][]{usoskin_Phi_05,herbst10}.
Thus, in order to compare our results with those of MB09, we repeat our computations based on Eq.~\ref{Eq:Q}
 to compute the global production $Q^*$ but using the same spectrum as MB09.
The result for $Q^*$ is shown by the grey curves in Fig.~\ref{Fig:MB}.
The overall agreement is within 5\% but $Q^*$ value is systematically higher than that of MB09.
The 5\% difference can be related to the slightly different numerical scheme and also to the fact
 that MB09 treated an $\alpha-$particle as four protons while we simulated them straightforwardly.
In addition, the way of considering the geomagnetic shielding by MB09 is simplified (scaling)
 compared to our consideration (direct computations).
We also compared the proton contributions (two dashed curves in Fig.~\ref{Fig:MB}) to $Q$ for the GCR spectrum discussed here (Eq.~\ref{Eq:ff})
 and that used in MB09.
The curves are nearly identical, suggesting that the difference in the used proton spectra is small and cannot
 be a cause for the observed systematic difference.
We however notice a great difference between the $\alpha-$ (and heavier) particle spectra used here and in MB09.
MB09 assumed of 12\% for $\alpha-$ and 1\% for heavier particle fraction in LIS (leading to $\approx$0.64 nucleonic ratio
 between heavier species to protons in GCR) basing on the data from \citet{simpson83}.
On the other hand, modern measurements (e.g., AMS, PAMELA) suggest that $\alpha-$particles above 10 GeV/nuc contribute 5--6\% (in
 particle number) to LIS of GCR
 leading to the nucleonic fraction of heavier species to protons of the order of 0.25-0.3
 outside the heliosphere \citep[e.g.][]{alcaraz00,alcaraz00a,adriani11,webber03,gaisser10},
 viz. half of that assumed by MB09.
Therefore, while we agree with MB09 in calculations of proton contribution into $Q$, they overestimate
 $^{14}$C production by heavier species of GCR, using outdated spectra.
This explains why the earlier results by MB99 and MB09 of $^{14}$C production are systematically higher than our present result.

\begin{figure}
\begin{center}
\resizebox{\hsize}{!}{\includegraphics{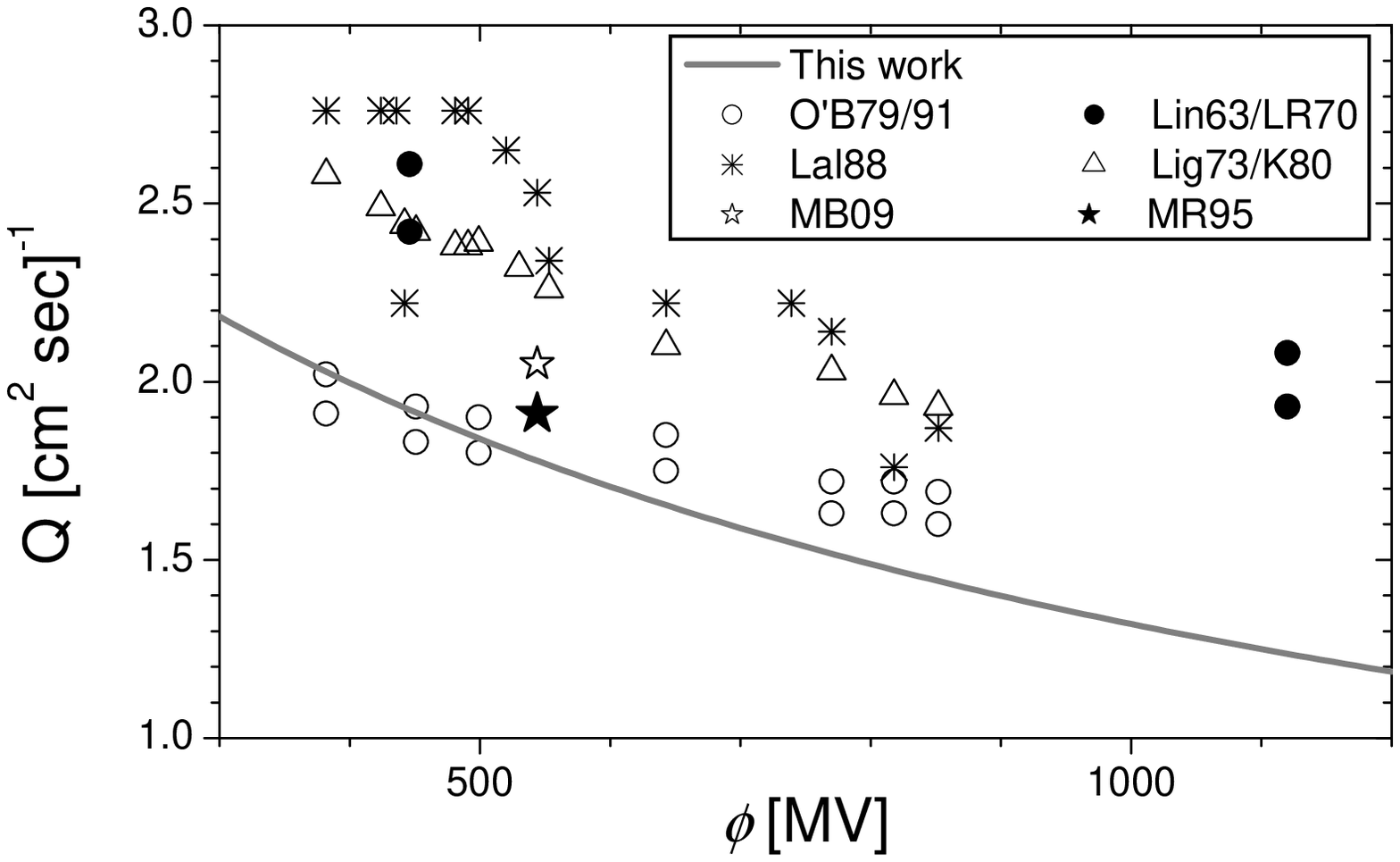}}
\end{center}
\caption{
Global $^{14}$C production as function of the modulation potential $\phi$ as defined in \citet{usoskin_bazi_11}.
The thick grey curve presents the present work's results.
Symbols corresponds to earlier works:
O'B79/91 \citep[Tab. 7 in][]{obrien79,obrien91};
Lin63/LR70 \citep[Tab. 1 in][]{lingenfelter63,lingenfelter70};
Lal88 \citep[Tabs. I and III in][]{lal88};
Lig73 \citep[Tab. 6 in][]{light73}, K80 \citep[Tab. 1 in][]{korff80};
MB09 \citep[Tab. 3 in][]{masarik09};
MR95 \citep[Tab. 1 in][]{masarik95}.
\label{Fig:Q14}}
\end{figure}

Next we compare predictions of our model with other models' results for specific periods of time as shown in Fig.~\ref{Fig:Q14}
 (exact data sets used are mentioned in the Figure caption).
One can see that our model predicts systematically lower production rates than most of other models, except of the
 model by \citet{obrien79} and \citet{obrien91}.
On the other hand, our yield function is generally consistent with others (Fig.~\ref{Fig:Y}), indicating
 that the difference must be related to the treatment of incoming GCR particle spectra and/or geomagnetic shielding
 and not to the atmospheric cascade simulations.
Models other than that by \citet{obrien79} were based on theoretical calculations and included outdated overestimated abundance of
 $\alpha-$particles, which explains the difference as discussed above.
Therefore, we conclude that our model more correctly calculates the $^{14}$C production as it agrees with
 the empirically-based models.

\section{$^{14}$C production by solar energetic particles}
\label{S:SEP}

We also calculated production of radiocarbon by solar energetic particles (SEP),
 because presently there is a wide range of the results \citep[e.g.,][]{lingenfelter70,usoskin_GRL_SCR06,hudson10,lavaletta11}.
Here we compute the expected production of $^{14}$C by the major known SEP events since 1951, using
 our calculated yield function (Table~\ref{Tab:Y}) and SEP event-integrated spectra as reconstructed by \citet{tylka09}.
The corresponding production rate is shown by big open dots in Fig.~\ref{Fig:50y} reduced to the monthly mean values.
One can see that only a few SEP events can produce significant enhancements in $^{14}$C production
 ($\approx 70$\% in the monthly mean for the SEP event of 23-Feb-1956, 40\% for 12-Nov-1960, 35\% for two events
 in Oct-1989 and $\approx$20\% for 29-Sep-1989).
However, when applied to the annual time scale (the standard tree-ring time resolution), it gives only a few percent effect
 for years of maximum solar activity and about 0.25\% of the total contribution over the considered period.
This is consistent with the earlier results by \citet{lingenfelter70} (1.1\% mean contribution of SEP into the
 global $^{14}$C production for 1954--1965, our model for the same period gives 0.8\%)
 and by \citet{usoskin_GRL_SCR06} (0.2\% for 1955--2005).
Note that MB09, however, gives much smaller value of 0.02\% for the SEP contribution to the global mean $^{14}$C production,
 which is probably caused by the neglect of the atmospheric cascade (and thus neutron capture channel) caused by SEPs
 \citep[cf.][]{masarik95}.

\section{Conclusions}
\label{S:disc}

\begin{itemize}
\item
We have performed full new calculation, based on a detailed Monte-Carlo simulation of the atmospheric cascade
 by a GEANT-4 tool PLANETOCOSMIC, of the $^{14}$C yield function.
 This is the first new calculation of the yield function since 1960-1970's, using modern techniques and methods,
  and the yield function is, for the first time ever, directly computed up to the energy of 1000 GeV/nuc
  (earlier models were limited to a few tens GeV/nuc and extrapolated to higher energies).
Our newly computed yield function gives the results which are in good agreement with \citet{obrien79} and
 consistent with most of the earlier models, within 10-20\%.

\item
We have calculated, using the new model and improved spectra of cosmic rays, the global production of
 $^{14}$C, which appears to be significantly lower than earlier estimates and closer to the values obtained
 from the carbon cycle inventory.
 The calculated modern global production rate is 1.64 atom/cm$^2$/sec, and the preindustrial rate (1750--1900 AD)
 is 1.88 atom/g/cm$^2$, which is essentially lower than earlier estimates of 2--2.5 atom/cm$^2$/sec.

\item
We explain that the earlier models (including a recent model by \citet{masarik09}) overestimate
 the contribution of $\alpha-$particle and heavier GCR species to the $^{14}$C production, because
 of the use of outdated spectra.

\item
We have calculated, on the basis of the new model, contribution to the global $^{14}$C production
 by SEP events, using updated energy spectra reconstructions by \citet{tylka09}.
 The mean contribution of the SEPs for the last 50 years is estimated to be $\approx$0.25\% of the total production.

\item
The present model provides an improved tool to calculate the $^{14}$C production in the Earth's atmosphere.
 Using the absolutely dated $^{14}$C calibration curve \citep{reimer_09}, one can reconstruct the variability
 of cosmic rays in the past \citep[e.g.,][]{solanki_Nat_04} which, along with other long-term solar proxies
 has applications to paleoastrophysics, paleomagnetism and paleoclimatology \citep[e.g.,][]{beer12}.

\end{itemize}

Supplementary materials related to this article can be found online at ...

\section*{Acknowledgements}
This work uses results obtained in research funded from the European Union's Seventh Framework Programme (FP7/2007-2013)
 under grant agreement No 262773 (SEPServer).
The High-Energy Division of Institute for Nuclear Research and Nuclear Energy - Bulgarian Academy of
 Sciences is acknowledged for the given computational time.
GAK was partly supported by the Program No. 22 of presidium RAS.
University of Oulu and the Academy of Finland are acknowledged for partial support.

\appendix
\section{Appendix: Details of numerical computations}
\label{App}
\setcounter{equation}{0}
\renewcommand{\theequation}{A\arabic{equation}}

Numerical computations were done using the GEANT-based Monte-Carlo simulation tool Planetocosmic \citep{desorgher05},
 which traces the atmospheric cascade induced by the primary cosmic ray particles in full detail,
 including the distribution of secondary particles.
 The Planetocosmic code has been recently verified \citep{usoskin_AG_09} to agree within $\approx 10$\% with
 another commonly used Monte-Carlo package CORSIKA \citep{heck98}, in the sense of energy deposition
 in the atmosphere.
The code simulates interactions and decays of various particles in the atmosphere in a wide range of energy.
For the computations, we applied a realistic spherical atmospheric model NRMLSISE-00 \citep{hedin91,picone02}.
The QGSP\_BIC\_HP hadron interaction model has been applied with the standard electromagnetic
 interaction model.

As an input for the simulations we used primary particles with fixed energy that impinge upon the
 top of the atmosphere at the random angle isotropically from the $2\pi$ solid angle.
All computations were normalized per one such simulated particle.
From the simulations we obtained the sum of secondary neutrons with energy within the $\Delta E$ energy bin centered at
 the energy $E_n$, crossing
 a given horizontal level (atmospheric depth $X$ g/cm$^2$), weighted with $|1/\cos{\theta}|$ (where $\theta$ is the zenith angle)
 to account for the geometrical factor, and divided by the energy bin width $\Delta E$.
This corresponds to the flux of secondary neutron with given energy $F(E_n,X)$ across a horizontal unit area, for the unit
 flux of primary cosmic rays on the top of the atmosphere.
On the other hand, for quasi-stationary flux of neutrons this can be expressed as 
\begin{equation}
F(E_n,X)\equiv n_n(E_n,X)\, v_n(E_n),
\end{equation}
where $n_n$ and $v_n$ are the concentration (in [MeV cm$^3$]$^{-1}$) and velocity of neutrons with energy $E_n$ at the atmospheric
 depth level $X$.
Let us denote  the integral columnar flux as
\begin{equation}
I(E_n)=\int_0^{X_m} F(E_n,X)\, dX,
\label{eq:I}
\end{equation}
where $X_m=1033$ g/cm$^2$ is the total thickness of the atmosphere.
Since our direct computations were performed down to energy of neutrons $E_1=$10 eV,
 we first computed the production of $^{14}$C by these super-thermal neutrons,
\begin{equation}
G_1=\sum_j{\int_{h}\left({\int_{E1}^\infty{F(E_n,X)\, n_j(h)\, \sigma_j(E_n)\, dE_n}}\right) dh},
\label{eq:Q1}
\end{equation}
where the outer integral is taken over the atmospheric height $h$, the concentration of target nuclei $n_j(h)$ is defined
 as a product of the air density $\rho$ and the
 content of the nuclei in a gram of air $\kappa_j$, $n_j(h)=\rho (h)\kappa_j$; $\sigma_j(E)$ is
 the cross-section of the corresponding reaction, and $dX=\rho(h)\, dh$, and summation is over
 target nuclei of different type (nitrogen $\kappa_{\rm N}=3.225\cdot 10^{22}$ atom/g; oxygen $\kappa_{\rm O}=8.672\cdot 10^{21}$
 atom/g; argon $\kappa_{\rm Ar}=1.94\cdot 10^{20}$ atom/g, we also accounted for the isotopic distribution within
 these groups).
Eqs.~\ref{eq:Q1} and \ref{eq:I} can be transformed so that
\begin{equation}
G_1=\sum_j{ \kappa_j {\int_{E1}^\infty{I(E_n)\, \sigma_j(E_n)\, dE_n}}},
\label{eq:Q1a}
\end{equation}
All the cross-sections, used here, have been adopted from the Experimental Nuclear Reaction Database (EXFOR/CSISRS)
 http://www.nndc.bnl.gov/exfor/exfor00.htm.

We note that all the processes related to leakage of neutrons from the atmosphere (to the space or to soil)
 as well as their decay are accounted for in the direct simulation.

Monte-Carlo simulations require extensive computational time in order to trace neutrons to thermal energy,
 thus compromising the statistical robustness of the results.
On the other hand, the fate of 10 eV neutrons can be easily modelled theoretically, because of the simplicity of
 the processes involved, which allows us to save computational time and improve accuracy of the computations.
The main process affecting epi-thermal neutrons in air is potential elastic scattering on N and O nuclei making
 neutrons to lose energy.
After each elastic scattering, a neutron has a uniform distribution of energy (in the laboratory frame) between
 its energy before the scattering $E_n$ and $\alpha\, E_n$ \citep[e.g., Chapter 7.2 in][]{fermi10}.
Here
\begin{equation}
\alpha={(A-1)^2\over (A+1)^2},
\end{equation}
where $A$ is the mass number of the target nucleus.
Then the probability for a neutron with the energy $E_n$ (if $E_1\leq E_n < E_1/\alpha$) before elastic
 scattering on a nuclei $j$ to have energy $E$ after the scattering so that $E<E_1$ is
 $(E_1-\alpha_j E_n)/(E_n(1-\alpha_j))$.
Accordingly the "flux" (in the energy domain) of neutrons crossing the energy boundary $E_1$ to (epi)thermal
 energies can be calculated as
\begin{equation}
N=\sum_j{\int_h{\int_{E_1}^{E_1/\alpha_j}{F(E_n,X)\, n_j(X)\, \sigma_{{\rm el,}j}(E_n){E_1-\alpha_j E_n\over E_n(1-\alpha_j)}dE_n\, dh}}}
\end{equation}
or, using Eq.~\ref{eq:I} as
\begin{equation}
N=\sum_j{\kappa_j   \int_{E_1}^{E_1/\alpha_j}{I(E_n) \sigma_{{\rm el,}j}(E_n){E_1-\alpha_j E_n\over E_n(1-\alpha_j)}dE_n}}
\end{equation}

Reactions involving neutrons are:
(1) N14(n,p)C14;  (2) O17(n,$\alpha$)C14; (3) N14(n,$\gamma$)N15; (4) O16(n,$\gamma$)O17;
 (5) O18(n,$\gamma$)O19 and (6) Ar40(n,$\gamma$)Ar41.
Note that only reactions (1) and (2) lead to production of $^{14}$C while others simply
 provide a sink for neutrons.
Cross-sections of neutron capture in all these reactions for energies below 10 eV can be
 expressed as
\begin{equation}
\sigma_j = {B_j\over v_n(E_n)},
\end{equation}
where $B_j$ is a constant.
Accordingly, the $^{14}$C production by these neutrons can be calculated as
\begin{equation}
G_2 = N\,\, {B_1\cdot\kappa_{\rm N14} + B_2\cdot\kappa_{\rm O17}\over \sum_j{B_j\cdot\kappa_j}}
\end{equation}
The bulk of radiocarbon $^{14}$C is produced via reaction (1) and about 0.001\%
 in reaction (2).
This is the main channel (95.8\%) of the neutron sink.
We have also considered leakage of neutrons from the upper atmospheric layers and decay
 of neutrons during their thermalization.
These processes appear to be unimportant.
In addition, we also computed possible contribution of secondary and primary protons to $^{14}$C
 production via spallation reactions (e.g., O16(p,X)C14).
These reactions are responsible for a negligible contribution to the total production.

Then the final production of $^{14}$C in the atmosphere by secondary neutrons corresponding to
 the primary cosmic ray particle with given energy is the sum of $G_1$ and $G_2$ and forms
 a point in the yield function $Y/\pi$.



\pagebreak
\renewcommand{\theequation}{\arabic{equation}}

\begin{table*}[t]
\caption{Normalized yield functions Y$_{\rm p}$/$\pi$ and Y$_{\alpha}$/$\pi$ of the atmospheric columnar $^{14}$C production (in atoms sr)
 by a nucleon of primary cosmic protons and $\alpha-$particles, respectively, with the energy given in GeV/nuc.
 For energy above 20 GeV/nuc, an $\alpha-$particle is considered to be identical to four protons.}
\begin{tabular}{lccccccccccccc}
\hline
E (GeV/nuc) & 0.1 & 0.3 & 0.5 & 0.7 & 1 & 3 & 7 & 10 & 19 & 49 & 99 & 499 & 999\\
\hline
proton & 0.025 & 0.26 & 0.72 & 1.29 & 2.07 & 5.19 & 8.32 & 9.72 & 12.40 & 17.45 & 23.24 & 48.30 & 72.73\\
$\alpha /4$ & 0.036 & 0.38 & 0.89 & 1.55 & 2.16 & 4.18 & 7.17 & 8.67 & 12.40 & 17.45 & 23.24 & 48.30 & 72.73\\
\hline
\end{tabular}
\label{Tab:Y}
\end{table*}

.
\pagebreak

\end{document}